\documentclass[aps,twocolumn,showpacs,preprintnumbers,amsmath,amssymb]{revtex4}
\usepackage{graphicx}
\usepackage{bm}

\begin{document}

\title{Symmetry of the order parameter in superconducting ZrZn$_2$}
\author{K. V. Samokhin}
\author{M. B. Walker} \affiliation{Department of Physics, University
of Toronto, Toronto, Ontario, Canada M5S 1A7}
\date{\today}

\begin{abstract}
We apply symmetry considerations to study the possible
superconducting order parameters in ferromagnetic ZrZn$_2$. We
predict that the presence and the location of the superconducting
gap nodes depend on the direction of magnetization $\bm{M}$.
In particular, if $\bm{M}$ is directed along the $z$ axis,
then the order parameter should always have zeros.
We also discuss how to determine the gap symmetry
in ZrZn$_2$ using ultrasound attenuation measurements.
\end{abstract}


\maketitle

Recently, superconductivity has been found in ZrZn$_2$ \protect\cite{pfl01},
which a weak itinerant ferromagnet.
The most surprising fact is that the superconductivity occurs only in the
ferromagnetic phase.
The exchange splitting of the Fermi surfaces makes a conventional singlet
BCS-like pairing strongly suppressed.
A number of theories have been proposed which show how the exchange by
spin fluctuations can lead to a triplet Cooper pairing both in the paramagnetic
and the ferromagnetic phases \cite{fay80}, or to the enhancement of the
superconducting critical temperature $T_c$ on the ferromagnetic side
\cite{kir01}.
Another feature of the phase diagram is that $T_c$ grows as pressure moves
away from the ferromagnetic quantum critical point, which can be
explained by the exchange-type interaction of the magnetic moments
of the Cooper pairs with the magnetization density \cite{walker02}.

Symmetry considerations can identify the possible order
parameters, even in the absence of a firmly established
microscopic mechanism of pairing, which is often the case for
unconventional superconductors. The presence of ferromagnetism
brings about a number of novel features in the symmetry analysis.
In this article, we give a detailed analysis of the
pairing symmetry in ZrZn$_2$ and discuss its consequences for the
superconducting gap structure. Our work has some overlap with the
recent theoretical studies \cite{mac01,fom01} of the gap
symmetry in another ferromagnetic superconductor UGe$_2$
\cite{sax00}. We find that the presence and location of the gap zeros
depend on the direction of magnetization. We also discuss in some detail
the design of ultrasonic attenuation experiments that can be used
for experimental probing the order parameter symmetry.

The symmetry group ${\cal G}$ of the system in the normal state is defined as
a group of transformations which leave the system Hamiltonian $H_0$ invariant,
i.e. $[G,H_0]=0$ for all elements $G\in{\cal G}$.
In non-magnetic superconductors, time reversal
symmetry $K$ is not broken,
and ${\cal G}=S\times K\times U(1)$, where $S$ is the space
group of the crystal, and $U(1)$ is the gauge group \cite{min99}.
In contrast, in magnetic superconductors,
time reversal symmetry is broken, and ${\cal G}=S_M\times U(1)$,
where $S_M$ is the magnetic space group which is a group
of symmetry operations leaving both
the crystal lattice (the microscopic charge density) and the magnetization
density $\bm{M}$ invariant \cite{lan60}. For example, if there is a crystal
point group rotation $R$ which transforms $\bm{M}$ to $-\bm{M}$,
then the combined operation $KR$ will be
an element of $S_M$, because the time reversal restores the original
$\bm{M}$ not affecting the lattice symmetry.
In the above expressions, it is assumed that the space group
elements act on the orbital and spin coordinates simultaneously, which is the
case when the spin-orbit coupling is present.
In the absence of spin-orbit coupling, the transformations of the orbital and
spin spaces are independent, so that ${\cal G}=S_{orb}\times G_{spin}\times 
U(1)$,
where $G_{spin}=SO(2)$ is the group of spin rotations about the direction of
$\bm{M}$.

The crystal structure of ZrZn$_2$ in the absence of ferromagnetism is described
by a face-centered cubic Bravais lattice, with the Zr atoms forming a
diamond structure. For cubic ferromagnets, the only two possibilities for the
easy direction of magnetization are a [001] or a [111] direction, and
these possibilities are analysed in the article.  Since a relatively small
magnetic field ($0.05$T at $T=1.75K$)
is required to line up the magnetic moments along a given
direction \cite{pfl01}, it is expected that experiments could be carried out for
$\bm{M}$ parallel to either [001] or [111].  The change in the
superconducting gap structure
when $\bm{M}$ is rotated by an external magnetic field is one of the
interesting and unusual properties of ZrZn$_2$ that could be investigated
experimentally.

If $\bm{M}$ is along [001], then $S_M$ is generated by: (i) the lattice
translations by the primitive vectors of the fcc lattice:
$\bm{t}_1=(a/2)(1,1,0)$,
$\bm{t}_2=(a/2)(0,1,1)$, and $\bm{t}_3=(a/2)(1,0,1)$, where $a$ is
the lattice constant;
(ii) the rotations $C_{4z}$ about the $z$ axis by an angle $\pi/2$ followed by
a fractional translation by a vector $\bm{\tau}=(a/4)(1,1,1)$;
(iii) the combined rotations $KC_{2x}$ about the $x$ axis by an angle $\pi$
accompanied by the time reversal; and (iv) the inversion $I$. The
point symmetry
of the crystal is described by the magnetic group
$\mathbf{D}_{4h}(\mathbf{C}_{4h})=\mathbf{D}_4(\mathbf{C}_4)\times\mathbf{C}_i$,
where $\mathbf{C}_i=\{E,I\}$. The subgroup
in parentheses (the unitary subgroup) incorporates all symmetry elements which
are not multiplied by the anti-unitary and anti-linear operation
$KC_{2x}$, i.e.
$\mathbf{D}_4(\mathbf{C}_4)=\mathbf{C}_4+KC_{2x}\times\mathbf{C}_4$.

If $\bm{M}$ is along [111], then $S_M$ is generated by: (i) the primitive
lattice translations, (ii) the rotations $C_{3xyz}$ about the [111]
direction by
an angle $2\pi/3$; (iii) the combined rotations $KC_{2\bar xy}$ about the
$[\bar 110]$ direction by an angle $\pi$ accompanied by the time reversal; and
(iv) the inversion $I$. The point symmetry is described by the magnetic group
$\mathbf{D}_{3d}(\mathbf{C}_{3i})=\mathbf{D}_3(\mathbf{C}_3)\times\mathbf{C}_i$,
where $\mathbf{D}_3(\mathbf{C}_3)=\mathbf{C}_3+KC_{2\bar xy}\times\mathbf{C}_3$.

Using the standard notation for the space group operations which combine
rotations $R$ and translations $\bm{t}$:
$\bm{r}\to(R|\bm{t})\bm{r}=R\bm{r}+\bm{t}$,
the transformation rules for the spinor wave functions can be written as
\begin{equation}
\label{Rdef}
     (R|\bm{t})\psi_\sigma(\bm{r})=
     [D^{(1/2)}(R)]_{\sigma\sigma'}\psi_{\sigma'}(R^{-1}(\bm{r}-\bm{t})).
\end{equation}
Here $\sigma=\uparrow,\downarrow$ is the spin projection on
the direction of $\bm{M}$, and
$D^{(1/2)}(R)$ is the spinor representation of rotations:
for a rotation by an angle $\theta$ around some axis
$\bm{n}$, $D^{(1/2)}(R)=U^{(s)}_{\bm{n}}(\theta)=
\exp(-i(\theta/2)(\bm{\sigma}\cdot\bm{n}))$.
We will also need the transformation rules under the time reversal operation:
\begin{equation}
\label{Kdef}
     K\psi_\sigma(\bm{r})=(i\sigma_2)_{\sigma\sigma'}\psi^*_{\sigma'}(\bm{r}),
\end{equation}
and the inversion:
\begin{equation}
\label{Idef}
     I\psi_\sigma(\bm{r})=\psi_{\sigma}(-\bm{r}).
\end{equation}

In the presence of the exchange field and spin-orbit coupling,
the single-particle wavefunctions are linear combinations
of the eigenstates of the spin
operator $s_z$: $\langle\bm{r}|\psi\rangle=u(\bm{r})|\uparrow\rangle+
v(\bm{r})|\downarrow\rangle$. Because the normal state Hamiltonian $H_0$ is
invariant with respect to the crystal lattice translations, the eigenfunctions
are the Bloch waves $\psi_{\bm{k}}(\bm{r})$ corresponding to wave vectors
$\bm{k}$ in the Brillouin zone. If the energy spectrum consists of
a single band which is doubly degenerate in zero
exchange field due to the Kramers theorem, then diagonalization
of the Hamiltonian in the presence of the exchange field results in
two non-degenerate energy bands
$\epsilon_\pm(\bm{k})$. The corresponding single-particle wave functions
have the form
\begin{equation}
\label{psi_pm}
  \langle\bm{r}|\bm{k},\pm\rangle=u^{(\pm)}_{\bm{k}}(\bm{r})|\uparrow\rangle+
    v^{(\pm)}_{\bm{k}}(\bm{r})|\downarrow\rangle.
\end{equation}
These states are referred to as the pseudospin states.
The operations from the group ${\cal G}$ conserve the pseudospin
in the following sense:
$G|\bm{k},\pm\rangle=\exp[i\phi_\pm(\bm{k},G)]|G_{orb}\bm{k},\pm\rangle$, with
$G_{orb}$ describing the ``orbital'' part of the symmetry operation, e.g.
rotations or reflections, and the undetermined phase factors $\phi_\pm$
coming from the freedom in choosing the overall phases of
$|\bm{k},\pm\rangle$ at every point of the Brillouin zone.
Here we adopt a convention introduced in Ref. \cite{ued85},
according to which the pseudospin states (\ref{psi_pm}) transform
similar to the
spin eigenstates $|\bm{k},\uparrow\rangle$ and $|\bm{k},\downarrow\rangle$.

Let us first consider the case $\bm{M}\parallel[001]$. From Eqs.
(\ref{Rdef},\ref{Kdef},\ref{Idef}), the transformation rules for
the creation operators of electrons in the states (\ref{psi_pm})
are:
\begin{equation}
\label{transforms}
\left.\begin{array}{ll}
   (C_{4z}|\bm{\tau}):& \ c^\dagger_{\bm{k},\pm}\to
                 e^{-i(C_{4z}\bm{k})\cdot\bm{\tau}}e^{\mp
i\pi/4}c^\dagger_{C_{4z}\bm{k},\pm} \\
   (KC_{2x}|\bm{0}):& \ \lambda c^\dagger_{\bm{k},\pm}\to
                 \pm i\lambda^*c^\dagger_{-C_{2x}\bm{k},\pm} \\
   I:& \ c^\dagger_{\bm{k},\pm}\to c^\dagger_{-\bm{k},\pm}.
\end{array}\right.
\end{equation}
Here $\lambda$ is an arbitrary $c$-number.

The pseudospin states can be used as a basis for constructing the
Hamiltonian which takes into account the Cooper pairing between
electrons with opposite momenta $\bm{k}$ and $-\bm{k}$.
Treating the Cooper interaction in the mean-field approximation, we
obtain $H=H_0+H_{sc}$, with the non-interacting part
\begin{equation}
\label{H_0}
     H_0=\sum\limits_{\bm{k}}
\Bigl[\epsilon_+(\bm{k})c^\dagger_{\bm{k}+}c_{\bm{k}+}+
     \epsilon_-(\bm{k})c^\dagger_{\bm{k}-}c_{\bm{k}-}\Bigr],
\end{equation}
describing two separate sheets of the Fermi surface corresponding to
different pseudospin indices, and
\begin{equation}
\label{H_sc}
     H_{sc}=\sum\limits_{\bm{k}}\sum\limits_{\alpha,\beta=\pm}
     \Bigl[\Delta_{\alpha\beta}(\bm{k})c^\dagger_{\bm{k}\alpha}
     c^\dagger_{-\bm{k},\beta}+\mathrm{h.c.}\Bigr].
\end{equation}
Here $\Delta_{++}(\bm{k})$ and $\Delta_{--}(\bm{k})$ represent 
the superconducting order parameters 
at the ``$+$'' and ``$-$'' sheets of the Fermi surface respectively, 
and $\Delta_{+-}(\bm{k})=-\Delta_{-+}(-\bm{k})$ is the
order parameter composed of quasiparticles on different sheets.
From the Pauli exclusion principle, $\Delta_{++}(\bm{k})$ and
$\Delta_{--}(\bm{k})$ are odd functions of $\bm{k}$, but
$\Delta_{+-}(\bm{k})$ does not have a definite parity. Separating
the odd and the even parts, the order parameter matrix can also be
cast in a more familiar form
$\Delta(\bm{k})=(i\bm{\sigma}\sigma_2)\bm{d}(\bm{k})+(i\sigma_2)\psi(\bm{k})$,
where $\bm{d}$ and $\psi$ are the pseudospin-triplet and the
pseudospin-singlet components respectively \cite{min99}. The fact
that the Fermi surface of ZrZn$_2$ consists of several sheets of
different topology \cite{maz97}, does not change our results.

From Eqs. (\ref{transforms}), the band spectra
$\epsilon_\pm(\bm{k})$ are invariant under the operations from the
point group $\mathbf{D}_{4h}$. Also, we obtain the transformation
rules for the order parameters under rotations $C_{4z}$:
\begin{eqnarray}
\label{C4}
     &&\Delta_{++}(\bm{k}) \to -i\Delta_{++}(C^{-1}_{4z}\bm{k})\nonumber\\
     &&\Delta_{--}(\bm{k}) \to +i\Delta_{--}(C^{-1}_{4z}\bm{k})\\
     &&\Delta_{+-}(\bm{k}) \to \Delta_{+-}(C^{-1}_{4z}\bm{k})\nonumber
\end{eqnarray}
(note the cancellation of the $\bm{\tau}$-dependent phase factors
on the right-hand side of these equations), and under the combined time
reversal and rotations $KC_{2x}$:
\begin{eqnarray}
\label{KC2x}
     &&\Delta_{++}(\bm{k}) \to \Delta^*_{++}(C^{-1}_{2x}\bm{k}) \nonumber\\
     &&\Delta_{--}(\bm{k}) \to \Delta^*_{--}(C^{-1}_{2x}\bm{k}) \\
     &&\Delta_{+-}(\bm{k}) \to \Delta^*_{+-}(-C^{-1}_{2x}\bm{k}).\nonumber
\end{eqnarray}

In the presence of the exchange band splitting $E_{ex}$,
the low-frequency part of the spectrum of excitations (e.g. spin fluctuations)
responsible for the interband Cooper pairing is cut out \cite{fay80}.
Since $E_{ex}$ is by far the largest energy scale in the system:
$E_{ex}\simeq 5\mathrm{mRy}\simeq 800\mathrm{K}$ \cite{santi01},
the pairing interactions
$c^\dagger_{\bm{k}+}c^\dagger_{-\bm{k},-}c_{\bm{k}'-}c_{-\bm{k}',+}$, which are
responsible for $\Delta_{+-}$, are negligibly small \cite{loff}
(some of the consequences of taking these interactions
into account will be discussed below).
On the other hand, the interband pairing terms
$c^\dagger_{\bm{k}+}c^\dagger_{-\bm{k},+}c_{\bm{k}'-}c_{-\bm{k}',-}$
can induce order parameters of the same symmetry on both sheets
of the Fermi surface. We expect the effect of these terms
to be small at small spin-orbit coupling, because they are absent at
zero spin-orbit coupling due to the spin conservation.

The superconducting order parameter which emerges at $T_c$ transforms according
to one of the irreducible representations $\Gamma$ of the normal state symmetry
group ${\cal G}$. It can be represented as the expansion
$\Delta_\Gamma(\bm{k})=\sum_i\eta_{\Gamma,i}f_i(\bm{k})$,
where $i$ labels the orbital basis functions, and $\eta_{\Gamma,i}$ are
the order parameter components
which enter, e.g., the Ginzburg-Landau free energy.
In our case, ${\cal G}$ contains the anti-unitary and anti-linear operation
$KC_{2x}$, and, instead of usual representations, one should use
co-representations of the magnetic point group $\mathbf{D}_4(\mathbf{C}_4)$,
which can be derived from one-dimensional representations
of the unitary subgroup $\mathbf{C}_4$ \cite{brad72}.
The results for odd co-representations are listed in Table \ref{table1}.
Note that the action of the unitary and anti-unitary
orbital symmetry elements on scalar functions $f(\bm{k})$ is defined as:
$Rf(\bm{k})=f(R^{-1}\bm{k})$, and $KRf(\bm{k})=f^*(-R^{-1}\bm{k})$.

\begin{table}
\caption{\label{table1} The character table and the examples
of the odd basis functions for the irreducible
co-representations of the magnetic point group
$\mathbf{D}_4(\mathbf{C}_4)$.
The overall phases of the basis functions are chosen so that
$KC_{2x}f_\Gamma(\bm{k})=f_\Gamma(\bm{k})$.
$\lambda_{1,2}$ are arbitrary real constants.}
\begin{ruledtabular}
\begin{tabular}{|c|c|c|c|}
   $\Gamma$  & $E$ & $C_{4z}$ &  $f_\Gamma(\bm{k})$ \\ \hline
   $A$      & 1   & 1  &  $k_z$  \\ \hline
   $B$      & 1   & $-1$ &  $k_z[\lambda_1(k_y+ik_x)^2+
            \lambda_2(k_y-ik_x)^2]$\\\hline
   $^1E$  & 1   & $i$ &  $k_y+ik_x$ \\ \hline
   $^2E$  & 1   & $-i$ &  $k_y-ik_x$
\end{tabular}
\end{ruledtabular}
\end{table}

If the superconductivity appears on the ``$+$''-sheet, then
the order parameter is $\Delta_{++}(\bm{k})$
[in terms of the vector order parameter
$\bm{d}(\bm{k})=d_z(\bm{k})\hat z+(d_+(\bm{k})(\hat x-i\hat y)
+d_-(\bm{k})(\hat x+i\hat y))/2$, it corresponds to $d_-=d_x-id_y$].
Using Eqs. (\ref{C4}) and Table \ref{table1},
we obtain the following expressions:
\begin{equation}
\label{delta_k}
\left.
\begin{array}{l}
  \Delta_{++,A}(\bm{k})=i\eta_{A}f_{^1E}(\bm{k})\\
  \Delta_{++,B}(\bm{k})=i\eta_{B}f_{^2E}(\bm{k})\\
  \Delta_{++,^1E}(\bm{k})=i\eta_{^1E}f_{B}(\bm{k})\\
  \Delta_{++,^2E}(\bm{k})=i\eta_{^2E}f_{A}(\bm{k}).
\end{array}
\right.
\end{equation}
The appearance of different representations on the left-hand and the
right-hand sides of these expressions can be easily understood if to look at
Eqs. (\ref{C4}). The transformed order parameter $\Delta_{++}$ has
an extra factor $-i$ which comes from the rotation of spin coordinates.
In terms of $\bm{d}(\bm{k})$, this factor
is the result of the rotation of the basis spin vector $\hat x+i\hat y$,
which transforms according to the $^2E$ representation.
Thus, for instance, the first line of Eqs. (\ref{delta_k}) follows from
the fact that $A={}^2E\times{}^1E$.
So far, we have discussed the transformation properties of the order
parameters (\ref{delta_k}) under the rotations from the unitary subgroup
$\mathbf{C}_4$. Because of our choice of the overall phase of the basis 
functions (see the caption to Table \ref{table1}) and the factors $i$
on the right-hand sides of Eqs. (\ref{delta_k}),
the effect of the anti-unitary operation $KC_{2x}$ on $\eta_\Gamma$ is
equivalent to complex conjugation: $KC_{2x}\eta_\Gamma=\eta^*_\Gamma$.

As seen from Eqs. (\ref{delta_k}) and Table \ref{table1},
the order parameters $\Delta_A$ and $\Delta_B$
vanish at the poles of the Fermi surface $k_x=k_y=0$, while the order
parameters
$\Delta_{^1E}$ and $\Delta_{^2E}$ vanish at the equator $k_z=0$. One can
prove that these gap zeros are not artifacts of our choice of the basis
functions but are imposed by symmetry.
Indeed, one of the elements of the unitary component $\mathbf{C}_{4h}$ of
the magnetic point group is the basal plane reflection
$\sigma_h=C_{2z}\times I$.
Therefore,
\begin{equation}
\label{sigma_h}
\sigma_hf_{A,B}(\bm{k})=f_{A,B}(k_x,k_y,-k_z)=-f_{A,B}(\bm{k}),
\end{equation}
so that $f_{A,B}(k_x,k_y,0)=0$, and 
$\Delta_{^1E}(k_x,k_y,0)=\Delta_{^2E}(k_x,k_y,0)=0$.
Similarly, under a four-fold rotation around the $z$ axis:
$$
C_{4z}f_{^1E,^2E}(\bm{k})=f_{^1E,^2E}(k_y,-k_x,k_z)=\pm
if_{^1E,^2E}(\bm{k}),
$$
hence $f_{^1E,^2E}(0,0,k_z)=0$, and 
$\Delta_{A}(0,0,k_z)=\Delta_{B}(0,0,k_z)=0$.
It also follows from Eq. (\ref{sigma_h}) that $f_{A}(\bm{k})$ and 
$f_{B}(\bm{k})$ go to zero
at $k_z=\pm\pi/a$, i.e. at the surface of the Brillouin zone, because
$(k_x,k_y,\pi/a)$ and $(k_x,k_y,-\pi/a)$ are equivalent points.
In order to take into account the crystal periodicity leading to
the presence of these additional gap zeros, one has to represent the basis
functions as the lattice Fourier series
$f(\bm{k})=\sum_nf_ne^{i\bm{k}\cdot\bm{R}_n}$,
where summation goes over the sites $\bm{R}_n$ of the Bravais lattice of the
crystal. The expansion appropriate for an odd order parameter
has the form
\begin{equation}
\label{lattice_series}
     f(\bm{k})=\sum\limits_n c_n\sin\bm{k}\cdot\bm{R}_n,
\end{equation}
where $\bm{R}_n$ are the sites of a fcc cubic lattice, which cannot
be transformed one into another by inversion.
In the nearest-neighbor approximation, we choose the following
set of $\bm{R}_n$'s: $\{\bm{R}_n\}=a/2\{(101),(\bar 101),(011),
  (0\bar 11),(110),(\bar 110)\}$.
Using the representation characters from Table \ref{table1}, we obtain
the basis functions which have symmetry-imposed zeros at the surface
of the Brillouin zone:
\begin{eqnarray*}
     &&f_A(\bm{k})=\sin\frac{k_za}{2}
         \left(\cos\frac{k_xa}{2}+\cos\frac{k_ya}{2}\right)\\
     &&f_B(\bm{k})=\sin\frac{k_za}{2}
         \left(\cos\frac{k_xa}{2}-\cos\frac{k_ya}{2}\right)\\
     &&f_{^1E}(\bm{k})=\cos\frac{k_za}{2}
         \left(\sin\frac{k_ya}{2}+i\sin\frac{k_xa}{2}\right)\\
     &&+\lambda_1\left[e^{\frac{i\pi}{4}}\sin\left(\frac{k_xa}{2}+\frac{k_ya}{2}
     \right)
     -e^{-\frac{i\pi}{4}}\sin\left(\frac{k_xa}{2}-\frac{k_ya}{2}\right)\right]\\
     &&f_{^2E}(\bm{k})=\cos\frac{k_za}{2}
         \left(\sin\frac{k_ya}{2}-i\sin\frac{k_xa}{2}\right)\\
     &&+\lambda_2\left[e^{-
\frac{i\pi}{4}}\sin\left(\frac{k_xa}{2}+\frac{k_ya}{2}\right)
     -e^{\frac{i\pi}{4}}\sin\left(\frac{k_xa}{2}-\frac{k_ya}{2}\right)\right].
\end{eqnarray*}
Here $\lambda_{1,2}$ are arbitrary real constants. The polynomial expressions
for the basis functions from Table \ref{table1} are recovered in the limit of
a ``small'' Fermi surface $\bm{k}\to 0$ [note that $f_B(\bm{k})$ from
Table \ref{table1} can be obtained by including the next-nearest-neighbors
in the expansion (\ref{lattice_series})].  It should be noted that
these nearest-neighbor results give also gap zeros not required by symmetry,
e.g. $f_B(\bm{k})=0$ on the plane $k_x=k_y$. These ``accidental'' zeros
will be removed if higher-neighbor terms are included, but if the
nearest-neighbor terms turn out to be dominant, experiment could find
indications of these accidental zeros.

The order parameter on the pseudospin-down sheet of the
Fermi surface is $\Delta_{--}(\bm{k})$
[in terms of $\bm{d}(\bm{k})$, it corresponds to
$d_+=d_x+id_y$, and the relevant basis spin vector
is $\hat x-i\hat y$, which transforms according to the $^1E$ representation].
Its $\bm{k}$-dependence is given by the following expressions:
\begin{equation}
\label{delta_kdown}
\left.
\begin{array}{l}
  \Delta_{--,A}(\bm{k})=i\eta_{A}f_{^2E}(\bm{k})\\
  \Delta_{--,B}(\bm{k})=i\eta_{B}f_{^1E}(\bm{k})\\
  \Delta_{--,^1E}(\bm{k})=i\eta_{^1E}f_{A}(\bm{k})\\
  \Delta_{--,^2E}(\bm{k})=i\eta_{^2E}f_{B}(\bm{k}).
\end{array}
\right.
\end{equation}

If we take into account
the interband pairing interaction of the form
$c^\dagger_{\bm{k}+}c^\dagger_{-\bm{k},+}c_{\bm{k}'-}c_{-\bm{k}',-}$,
then both $\Delta_{++}$ and $\Delta_{--}$ are non-zero and correspond
to the same irreducible co-representation of the magnetic point group.
Comparing Eqs. (\ref{delta_k}) and (\ref{delta_kdown}), we see that,
although the orbital symmetries of
$\Delta_{++}$ and $\Delta_{--}$ are different, they have
the symmetry-imposed gap nodes at the same locations on both sheets
of the Fermi surface. For example, if
the order parameter corresponds to the $A$
co-representation, then both
$\Delta_{++,A}(\bm{k})\sim f_{^1E}(\bm{k})$ and
$\Delta_{--,A}(\bm{k})\sim f_{^2E}(\bm{k})$ have
point nodes at $k_x=k_y=0$.

The gap nodes disappear only if the interband pairing interactions
$c^\dagger_{\bm{k}+}c^\dagger_{-\bm{k},-}c_{\bm{k}'-}c_{-\bm{k}',+}$ are
taken into account. These terms induce a non-zero order parameter
$\Delta_{+-}$, whose momentum dependence in the triplet channel,
according to Eqs. (\ref{C4}) and Table \ref{table1}, is given by
$\Delta_{+-,\Gamma}(\bm{k})\sim f_\Gamma(\bm{k})$, where 
$\Gamma=A,B,{}^1E$, or ${}^2E$.
To see explicitly how the structure of the nodes in the different
components of the gap function is translated into zeros of the
spectrum of Bogoliubov quasiparticle excitations, it is necessary
to diagonalize the Hamiltonian of Eqs.\ (\ref{H_0}) and
(\ref{H_sc}). This gives the following condition for the energy 
$E({\bm{k}})$ of a quasiparticle to be zero at some $\bm{k}$:
\begin{eqnarray}
\label{EB}
    &&\epsilon_+^2\epsilon_-^2+\epsilon_+^2|\Delta_{--}|^2 +
        \epsilon_-^2|\Delta_{++}|^2   \nonumber \\
    &&+\epsilon_+\epsilon_-(|\Delta_{+-}|^2+|\Delta_{+-}|^2)+ 
    |\mathrm{det}\;\Delta|^2 = 0.
\end{eqnarray}
The condition for zeros in the excitation energy 
on the ``+''-sheet of the
Fermi surface (i.e. at $\epsilon_+(\bm{k})=0$) is that 
$\Delta_{++}(\bm{k})=\Delta_{+-}(\bm{k})=0$, while for
zeros in the excitation energy on the ``$-$''-sheet 
(i.e. at $\epsilon_-(\bm{k})=0$)
we must have $\Delta_{--}(\bm{k})=\Delta_{+-}(\bm{k})=0$. Thus a
nonzero $\Delta_{+-}$ will remove the nodes in the spectrum of
elementary excitations. For example, $\Delta_{+-,A}(\bm{k})
\sim f_A(\bm{k})$ does 
not vanish at $k_x=k_y=0$, so that these point nodes should be filled. 
However, as discussed above, this effect 
is expected to be negligibly small.

In the absence of a complete understanding of the microscopic mechanism
of the superconductivity in ZrZn$_2$,
one cannot tell which order parameter from the lists
(\ref{delta_k}) and (\ref{delta_kdown}) corresponds
to the highest critical temperature. For example, if the superconductivity is due
to the exchange by spin fluctuations, then, at vanishing spin-orbit coupling,
the order parameters $\Delta_{++,A},\Delta_{++,B}$, and $\Delta_{++,{}^2E}$
correspond to the $p$-wave equal-spin-pairing superconducting states
studied in Ref. \cite{fay80}; in these terms,
$\Delta_{++,{}^1E}$ corresponds to $f$-wave pairing.
In Ref. \cite{walker02}, a simple phenomenological model of the phase diagram
of ZrZn$_2$ was proposed. The basic idea was that the underlying order
parameter is a vector quantity transforming according to a three-dimensional
representation of the cubic group. Then, the exchange-type interaction of the
magnetic moments of Cooper pairs with the ferromagnetic magnetization
splits the superconducting critical temperature and lowers the
dimensionality of the order parameter from three to one. In this model,
the order parameters $\Delta_{^1E}$ and $\Delta_{^2E}$ are the possible
ones, and the experimental determination of the gap symmetry will thus be
helpful in assessing the validity of the model.

A similar analysis can be done if the easy axis for magnetization is [111].
In this case, the band spectra are invariant under the operations from the
group $\mathbf{D}_{3d}$, the relevant magnetic point group is
$\mathbf{D}_3(\mathbf{C}_3)$, and the transformation rules (\ref{C4}) and
(\ref{KC2x}) for the order parameter are replaced by
\begin{eqnarray}
\label{C3}
     &&\Delta_{++}(\bm{k}) \to e^{-2i\pi/3}\Delta_{++}(C^{-1}_{3xyz}\bm{k})
     \nonumber\\
     &&\Delta_{--}(\bm{k}) \to e^{+2i\pi/3}\Delta_{--}(C^{-1}_{3xyz}\bm{k})\\
     &&\Delta_{+-}(\bm{k}) \to \Delta_{+-}(C^{-1}_{3xyz}\bm{k}),\nonumber
\end{eqnarray}
under rotations $C_{3xyz}$, and
\begin{eqnarray}
\label{KC2xy}
     &&\Delta_{++}(\bm{k}) \to \Delta^*_{++}(C^{-1}_{2\bar xy}\bm{k}) \nonumber\\
     &&\Delta_{--}(\bm{k}) \to \Delta^*_{--}(C^{-1}_{2\bar xy}\bm{k}) \\
     &&\Delta_{+-}(\bm{k}) \to \Delta^*_{+-}(-C^{-1}_{2\bar xy}\bm{k}),\nonumber
\end{eqnarray}
under the combined operation $KC_{2\bar xy}$.

\begin{table}
\caption{\label{table2} The character table and the examples
of the odd basis functions for the irreducible
co-representations of the magnetic point group
$\mathbf{D}_3(\mathbf{C}_3)$, $\omega=e^{2\pi i/3}$.
The overall phases of the basis functions are chosen so that
$KC_{2\bar xy}f_\Gamma(\bm{k})=f_\Gamma(\bm{k})$.}
\begin{ruledtabular}
\begin{tabular}{|c|c|c|c|c|}
   $\Gamma$  & $E$ & $C_{3xyz}$ & $C^{-1}_{3xyz}$ & $f_\Gamma(\bm{k})$ \\ \hline
   $A$      & 1   & 1 & 1  &  $k_x+k_y+k_z$  \\ \hline
   $^1E$  & 1   & $\omega$ & $\omega^*$ &
                 $e^{-i\pi/3}k_x-k_y+e^{i\pi/3}k_z$ \\ \hline
   $^2E$  & 1   & $\omega^*$ & $\omega$  &
                 $e^{i\pi/3}k_x-k_y+e^{-i\pi/3}k_z$
\end{tabular}
\end{ruledtabular}
\end{table}

Using Table \ref{table2}, we obtain the following 
$\bm{k}$-dependences of the order parameter at the pseudospin-up sheet:
\begin{equation}
\label{delta_k111}
\left.\begin{array}{l}
  \Delta_{++,A}(\bm{k})=i\eta_{A}f_{^1E}(\bm{k})\\
  \Delta_{++,^1E}(\bm{k})=i\eta_{^1E}f_{^2E}(\bm{k})\\
  \Delta_{++,^2E}(\bm{k})=i\eta_{^2E}f_{A}(\bm{k}).
\end{array}\right.
\end{equation}
As above, the factors $i$ here guarantee that the action of $KC_{2\bar xy}$
on $\eta_\Gamma$ is equivalent to complex conjugation.

A consequence of the above results is that
there are no gap nodes
for $\Delta_{^2E}$, whereas $\Delta_A$ and $\Delta_{^1E}$ have
point nodes where the line $k_x = k_y = k_z$ cuts the Fermi surface.
For completeness, we also give the expressions for the basis functions
of the magnetic point group $\mathbf{D}_3(\mathbf{C}_3)$ in terms of the lattice
Fourier series in the nearest-neighbor approximation:
\begin{eqnarray*}
     f_A(\bm{k})&=&S^{+}_1+S^{+}_2+S^{+}_3\\
     &&+i\lambda_1\left(S^{-}_1+S^{-}_2+S^{-}_3\right)\\
     f_{^1E}(\bm{k})&=&\omega^*S^{+}_1+\omega S^{+}_2+S^{+}_3\\
     &&+i\lambda_2\left(\omega^*S^{-}_1+\omega S^{-}_2+S^{-}_3\right)\\
     f_{^2E}(\bm{k})&=&\omega S^{+}_1+\omega^*S^{+}_2+S^{+}_3\\
     &&+i\lambda_3\left(\omega S^{-}_1+\omega^*S^{-}_2+S^{-}_3\right),
\end{eqnarray*}
where $S^{\pm}_1=\sin(k_xa/2\pm k_ya/2)$, $S^{\pm}_2=\sin(k_ya/2\pm k_za/2)$,
$S^{\pm}_3=\sin(k_za/2\pm k_xa/2)$, and $\lambda_{1,2,3}$ are arbitrary
real constants.

For the order parameter at the ``$-$''-sheet of the Fermi surface,
\begin{equation}
\label{delta_k111down}
\left.\begin{array}{l}
  \Delta_{--,A}(\bm{k})=i\eta_{A}f_{^2E}(\bm{k})\\
  \Delta_{--,^1E}(\bm{k})=i\eta_{^1E}f_{A}(\bm{k})\\
  \Delta_{--,^2E}(\bm{k})=i\eta_{^2E}f_{^1E}(\bm{k}).
\end{array}\right.
\end{equation}
There are no gap nodes for $\Delta_{^1E}$, but $\Delta_A$
and $\Delta_{^2E}$ have point nodes at $k_x = k_y = k_z$.
If the interband hybridization of the form
$c^\dagger_{\bm{k}+}c^\dagger_{-\bm{k},+}c_{\bm{k}'-}c_{-\bm{k}',-}$
is taken into account, the order parameters are non-zero on both
sheets of the Fermi surface. From Eqs. (\ref{delta_k111}) and
(\ref{delta_k111down}), we see that both
$\Delta_{++,A}(\bm{k})\sim f_{^1E}(\bm{k})$ and
$\Delta_{--,A}(\bm{k})\sim f_{^2E}(\bm{k})$ vanish on the line
$k_x = k_y = k_z$. The gap nodes
disappear only in the presence of the interband pairing
$c^\dagger_{\bm{k}+}c^\dagger_{-\bm{k},-}c_{\bm{k}'-}c_{-\bm{k}',+}$,
which induces the order parameter 
$\Delta_{+-,\Gamma}(\bm{k})\sim f_\Gamma(\bm{k})$, where 
$\Gamma=A,{}^1E$, or ${}^2E$.
Again, we expect $\Delta_{+-}$ to be negligibly small in the presence
of the large exchange field.

The presence of the gap nodes would manifest themselves in
power-law temperature dependences of the thermodynamic and
transport properties \cite{min99}. For example, the electronic
specific heat at low temperatures should be
$C(T)/T=\gamma_0+\gamma_1T$ for the line nodes, and
$C(T)/T=\gamma_0+\gamma_1T^2$ for the point nodes. The
temperature-independent contributions on the right hand side of
these equations come from the normal excitations at the unpaired
sheet of the Fermi surface. If the magnitudes of both
order parameters $\Delta_{++}$ are $\Delta_{--}$ are comparable
(we expect this to be the case only if the spin-orbit coupling
is strong enough), then $\gamma_0$ is absent and a power-law dependence
should be observed.

One of the most powerful methods of determining the presence and
the location on the Fermi surface of gap nodes in unconventional
superconductors has been ultrasonic attenuation
\cite{shi86,ell96,mor96,lup01,wal01}. The method is based on finding which
sound waves are particulary weakly attenuated by the nodal
quasiparticles.  Nodal quasiparticles are ``inactive'' in
attenuating a particular sound wave if the electron-phonon
interaction for the nodal quasiparticle with the particular sound
wave is zero. Symmetry arguments determining the inactive nodes
have been developed in a previous article \cite{wal01}. We refer
to that article for a detailed treatment of the basic ideas, and
give here the extension of the arguments which is necessary for
treating magnetic groups.

If a symmetry operation of the crystal (i.e. an element of its
magnetic point group) leaves the wave vector ${\bm k}$ characterizing a
given electron state invariant, then the interaction of this
electron with certain phonons can be shown to be zero. Consider a
phonon of wavevector ${\bm q}$ and polarization direction
${\bm e}$. The interaction of the given electron with the given phonon
can be shown to be zero if the symmetry operation causes
an odd number of changes of sign of the two quantities
$i\bm{q}$ and $\bm{e}$ (the factor $i$ is important because the time
reversal operation contains complex conjugation).
The transformation rules for ${\bm k}$, ${\bm q}$, and ${\bm e}$ are:
(i) under the point-group operations $R$, they transform like polar vectors,
i.e. ${\bm k}\to R^{-1}{\bm k}$, {\em etc}.; and (ii) under the
combined operations $KR$, $\bm{k}\to-R^{-1}\bm{k}$, $\bm{e}\to R^{-1}\bm{e}$,
and $\bm{q}\to-R^{-1}\bm{q}$, so that $i\bm{q}\to iR^{-1}\bm{q}$.
For example, suppose that the magnetic
group contains the symmetry element $KC_{2x}$.  The wave
vectors ${\bm k}$ of electrons lying in the $k_x=0$ plane are
invariant under $KC_{2x}$. According to the rule just stated,
these electrons have zero interaction with transverse phonons
having their wave vectors along the $x$ axis
because, under the operation $KC_{2x}$, $i{\bm q}$
remains invariant, but ${\bm e}$ changes sign.

As shown above, one should expect the order parameter in ZrZn$_2$
to have nodes if $\bm{M}\parallel[001]$, when the magnetic
point group is $\mathbf{D}_4(\mathbf{C}_4)$. The gap nodes are always active
for longitudinal sound waves.
If the order parameter has point nodes
($\Delta_{A}$ or $\Delta_{B}$ in Eqs. (\ref{delta_k})),
then these point nodes are inactive for the transverse waves
T100 and T110 polarized either in the basal plane or
along [001], and also for
the waves T001 polarized either along [100] or [110]. (By definition a
Thkl sound wave is a transverse wave having its wave vector $\bm{q}$
along the [hkl] direction.)
If the order parameter has line nodes in the plane $k_z=0$
($\Delta_{^1E}$ or $\Delta_{^2E}$ in Eqs. (\ref{delta_k})),
then these line nodes are inactive for the transverse waves T100 and T110
polarized along [001], and also for T001 waves  polarized
either along [100] or [110]. Note that the attenuation of the
T100 and T110 waves polarized in the basal plane can be used
to distinguish between the equatorial line nodes and the point nodes,
because the former are active, but the latter are inactive.
The presence of unpaired electrons on one of the sheets of the Fermi
surface will not cause difficulties in symmetry determination by
ultrasonic attenuation as this will simply make a contribution to
the low-temperature temperature-independent background, which is
easily distinguished from the temperature-dependent contribution
of the gapped sheet (or sheets) of the Fermi surface.

To summarize, we have studied the symmetry of the superconducting
order parameter in ZrZn$_2$. If the spin-orbit coupling is weak then
superconductivity should appear on only one of the sheets of the Fermi surface.
The interband scattering can, in principle, induce non-zero
order parameters on other sheets and also fill the gap zeros,
but we expect these effects to be small. The symmetry of the order parameter
depends on the direction of the easy axis for magnetization.
If $\bm{M}\parallel[001]$, then the magnetic point group is 
$\mathbf{D}_4(\mathbf{C}_4)$,
and the order parameter goes to
zero on the line $k_x=k_y=0$ for the gap symmetries
$A$ and $B$, or on the planes $k_z=0,\pm\pi/a$ for the symmetries $^1E$ and 
$^2E$, on both sheets of the Fermi surface.
The positions of the gap zeros can be probed by ultrasonic attenuation
measurements, and to assist in the design of appropriate experiments
we have given a detailed discussion of the
zeros of the electron-phonon interaction in ferromagnetic ZrZn$_2$ which are
imposed by the magnetic point symmetry.
If $\bm{M}\parallel[111]$, then the magnetic point
group is $\mathbf{D}_3(\mathbf{C}_3)$,
and the order parameter has point zeros on the line $k_x=k_y=k_z$ 
on both sheets of the Fermi surface, for the gap symmetry $A$,
and on one of the sheets, for the symmetries $^1E$ and $^2E$.
It should be possible to fix the magnetization density $\bm{M}$ along an
arbitrary crystallographic direction by the application of an external
magnetic field, and hence to determine the gap structure for ZrZn$_2$ for
$\bm{M}$ along both [001] and [111] and to find the changes that occur
when $\bm{M}$ is rotated from [001] to [111].

We acknowledge a stimulating discussion with Louis Taillefer and
support from the Canadian Institute for Advanced Research and from
the Natural Sciences and Engineering Research Council of Canada.


\begin{thebibliography}{99}
\bibitem{loff}
We also neglect here the possibility of the Cooper pairs having a non-zero
momentum, i.e.
$\langle c^\dagger_{\bm{k}+\bm{q},+}c^\dagger_{-\bm{k},-}\rangle\neq 0$
(Larkin-Ovchinnikov-Fulde-Ferrell state).
\end{thebibliography}
\end{document}